\documentstyle[11pt,newpasp]{article}

\begin{document}

\title{Gas dynamics and disk formation in 3:1 mergers}
\author{Thorsten Naab \& Andreas Burkert}
\affil{Max-Planck-Institut f\"ur Astronomie, K\"onigstuhl 17, 69117
Heidelberg}

\begin{abstract}
We performed N-body/SPH simulations of merging gas rich disk-galaxies
with a mass ratio of 3:1. A stellar disk and bulge component and a
dark halo is realized with collisionless particles, the gas is
simulated using SPH particles. In this paper we focus on the gas
dynamics in mergers and its influence on the structure of the final
merger remnant, neglecting star formation. During the merger around
$8\times 10^9 M_{\odot}$ of gas accumulate in the central region of
the merger remnant (inside 300 pc). Part of the gas falls to the
center, channeled by a tidally induced bar in the more massive
galaxy. A peak accretion rate of 150 solar masses/year is reached
when the two galaxy centers merge. It is likely that the gas will
experience a central starburst and/or fuel a central black hole. The
gas in the outer regions accumulates in dense knots within tidal tails
which could lead to the formation of open clusters or dwarf
satellites. Later on, the gas knots loose angular momentum by 
dynamical friction and also sink to the center of the remnant, thereby
increasing the central gas content. In the end, the gas settles in a
central power-law disk with surface density $\Sigma \propto r^{-2}$
surrounded by dense, orbiting gas clumps. The stellar profile is of de
Vaucouleurs-type, like in simulations of pure stellar mergers    
\end{abstract}


\section{Initial Conditions and merger model}
The disk-galaxies are constructed in dynamical equilibrium (Hernquist,
1993) and consist of an exponential stellar and gaseous disk of the
same mass, a bulge with a Hernquist-profile, and a pseudo-isothermal
dark halo. The small galaxy has 1/3 of the mass and 1/3 of the
particles of the massive one. The scale length is reduced by
$(1/3)^{1/2}$. In total we used 88888 collisionless particles and
26666 gas particles simulated using SPH with an isothermal equation
of state. Both galaxies approach each other on nearly parabolic
prograde orbits with slightly inclined disks relative to the orbital
plane. The simulations were performed on a Sun ULTRA 60 workstation.
 
\section{Results}
The hypothesis that pure stellar 3:1 mergers of disk galaxies lead to
fast rotating and isotropic (Barnes, 1998), disky galaxies (Naab,
Burkert \& Hernquist, 1999) is tested with a model that also follows
the evolution of a massive gas component. If only the stellar part is
taken into account the remnant still follows a  $r^{1/4}$ profile
(Figure 1). It has predominantly disky
isophotes, isotropic velocity dispersion and a small amount of
minor-axis rotation, although the amount of diskiness is significantly
reduced and the remnant has a much rounder shape.    
In 1:1 mergers most of the gas accumulates  in a central gas knot
(Barnes \& Hernquist, 1996; Mihos \& Hernquist, 1996). In contrast,
the gas in 3:1 mergers has much larger specific angular momentum. It
therefore settles into an extended inner disk component (Figure
1). Simulations like this can help to understand the formation of disky
elliptical galaxies or the formation of central power-law or disk-like 
structures that are observed in fast rotating elliptical
galaxies (Faber et al., 1997). The detailed influence of gas on the
global structure and formation of elliptical galaxies is not well
understood and might depend strongly on the star formation history
which has been excluded in the present simulations.    

\begin{figure*}
\includegraphics{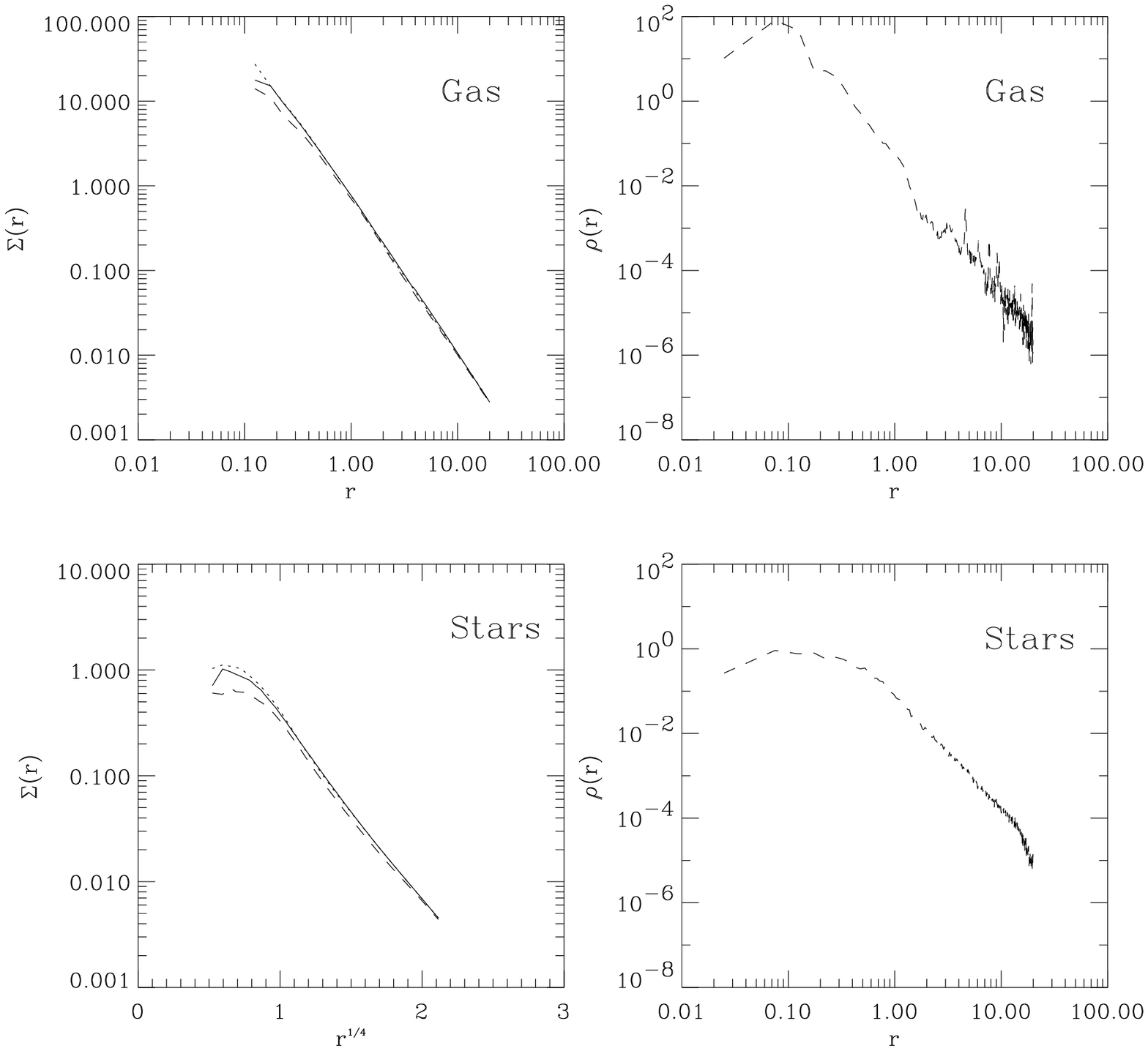}
\includegraphics{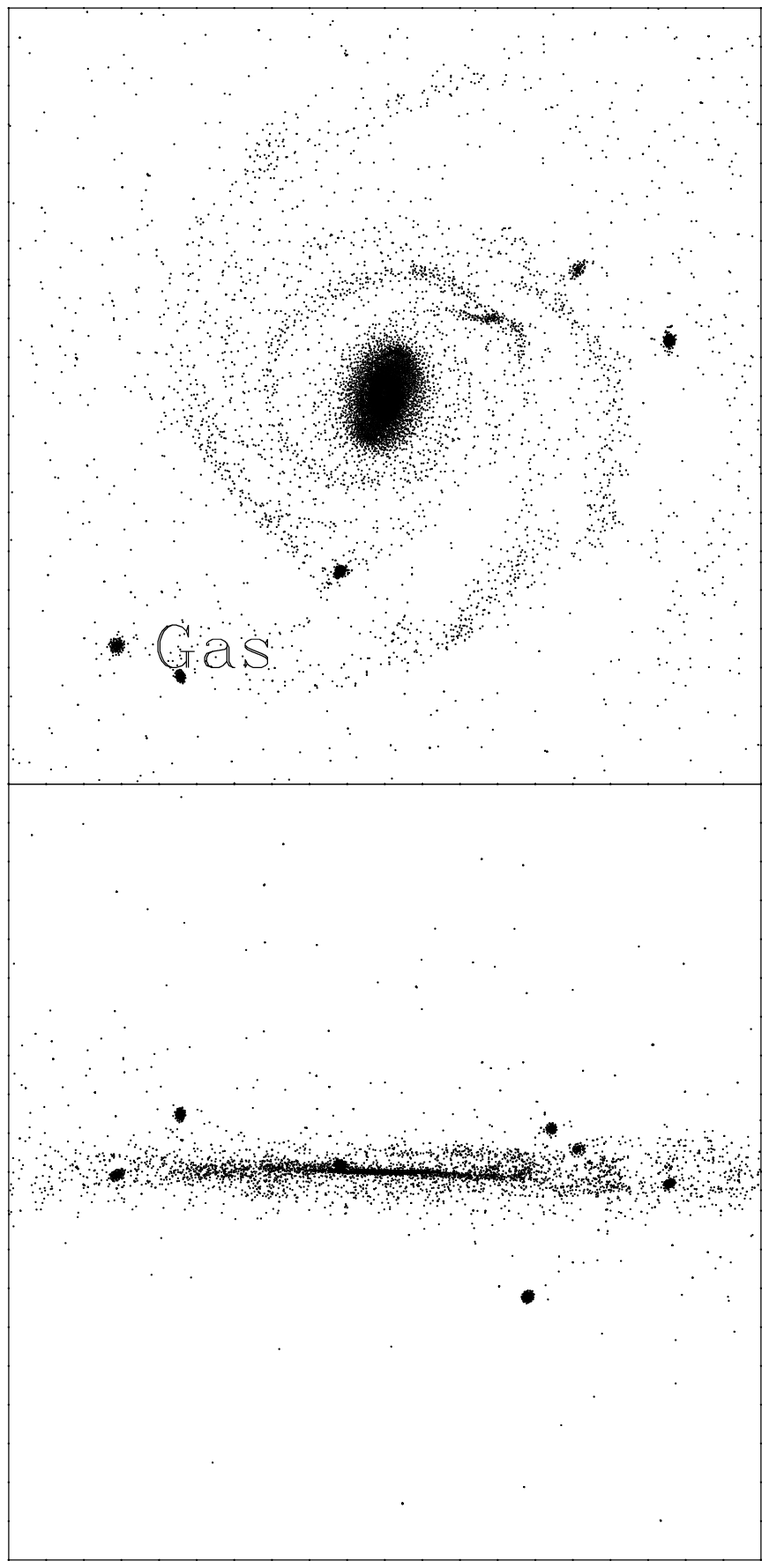}
\vspace{7cm}
\caption{{\it Left:} Surface density and spatial density of the gas
and the stars at the end of the simulation. {\it Right:} Distribution of
gas particles of the disk formed at the end of the simulation seen
face on (upper panel) and edge on (lower panel). The boxlength
corresponds to 10 kpc, if the size of the initial disk is scaled to the
Milky Way.} 
\end{figure*}

\end{document}